\documentclass[12pt]{article}
\usepackage{graphicx,amsmath,lineno}
\usepackage{color}

\newlength{\dinwidth}
\newlength{\dinmargin}
\begin{document}

\def\be{\begin{equation}}                                                    

\def\lapproxeq{\lower .7ex\hbox{$\;\stackrel{\textstyle                                                    
<}{\sim}\;$}}                                                    
\def\gapproxeq{\lower .7ex\hbox{$\;\stackrel{\textstyle                                                    
>}{\sim}\;$}}                                                    
\def\be{\begin{equation}}                                                    
\def\ee{\end{equation}}                                                    
\def\bea{\begin{eqnarray}}                                                    
\def\eea{\end{eqnarray}}

\begin{flushright}                                                    
IPPP/19/16\\
\today \\                                                    
\end{flushright} 

\vspace{1cm}

\begin{center}
{\Large\bf BFKL Pomeron and the survival factor\footnote{A contribution to the memorial volume entitled `From the past to the future - the legacy of Lev Lipatov' edited by J. Bartels {\it et al}~.}}\\
\vspace{0.5cm}
 V.A. Khoze, A.D. Martin, M.G. Ryskin\\
 \vspace{1cm}

\begin{abstract} 
We consider the absorptive corrections and the rapidity gap survival factor which are necessary to provide the unitarization of the BFKL Pomeron. In particular we discuss the role of the enhanced screening diagrams.
\end{abstract}
\end{center}
\vspace{1.5cm}
\section{Introduction}

Perhaps we may start with a short, but heartfelt, tribute to Lev Lipatov.  We were very fortunate to enjoy long-term scientific interactions with Lev; it was a real privilege to collaborate with him on various projects.  All his research was at a deep theoretical level and he had a very clear vision of his objectives.  His insight was profound and legendary.  He seemed materially devoid of ambition and yet scientifically his ambition for the truth was enormous. His deep understanding enabled him to extend our view of nature in many ways.  The ten or so years that Lev Lipatov and Victor Fadin spent meticulously extending BFKL to next-to-leading order were a remarkable achievement.

As a colleague and friend Lev was humble and considerate. He was always willing and patient to explain his ideas to those less clever than himself. Indeed, Lev never refused to discuss a topic in depth and took care of the young theorists in the Theory Division of the St. Petersburg Nuclear Physics Institute, creating what became known world-wide as the Lipatov School. We miss him greatly as a person and as a scientist.

An important aspect of Lev Nikolaevich Lipatov's prolific and penetrating research career concerned the application of perturbative Quantum Chromodynamics (QCD) to  
high-energy interaction amplitudes.  In particular how to describe the proton's behaviour in high-energy collisions, a topic extremely important to the analysis of data from the particle physics experiments around the world. In fact, Lipatov was the unique co-author to both of the two relevant theoretical breakthroughs: the DGLAP \cite{DGLAP} and BFKL \cite{BFKL} formalisms.  This memorial article concerns one of the consequences of the BFKL formalism. It deals with the importance of absorptive corrections and informs us about the special behaviour of the survival factors of the Large Rapidity Gaps (LRGs) which can occur in high-energy proton-proton scattering.

In the BFKL formalism, appropriate to the small $x$ domain, high-energy $pp$
scattering is, to leading ($\alpha_s\cdot\ln s )^n$ order, driven by Pomeron exchange, which at
the parton level is composed of reggeized two-gluon exchange.
A crucial feature of the BFKL Pomeron is the fact that
within perturbative QCD it leads to the power growth of high-energy cross
section with energy, $\sqrt s$. That is, the high-energy (proton-proton) forward scattering amplitude, $A(s,t=0)$, increases with energy as
\begin{equation}
\label{a1}
A_{\rm BFKL}(s,t=0)~=~i\alpha^2_s \left(\frac{s}{s_0}\right)^{\alpha_P(0)}
\end{equation}
where $\alpha_s$ is the QCD coupling and the Pomeron intercept $\alpha_P (0)$ is such that
\be
\alpha_P(0)~\equiv~1+\omega_0~>~1.
\ee
 In the leading Log approximation $\omega_0=(N_c\alpha_s/\pi)4\ln 2$~\cite{BFKL}. If we account for the next-to-leading corrections then $\omega_0\simeq 0.2-0.3$~\cite{NLL}.

Recall that this result was obtained for a very small $\alpha_s$
\be
\alpha(s)\ll
1 ~~~~{\rm but}~~~~ \alpha_s\ln(s/s_0)\sim O(1).
\ee
 At larger energies, when the value of $Y=\ln(s/s_0)$ increases, 
  the amplitude
(\ref{a1}) starts to violate the Froissart bound~\cite{Fr}. However, the power growth of the amplitude ($s^{\omega_0+1})$ is compensated  by {\it absorptive corrections} described by the multi-Pomeron diagrams shown in Fig.1.  Actually the corrections become large already for
\be
 \omega_0 Y~>~2\ln(1/\alpha_s)
 \ee

At a fixed impact parameter, $b$, sum of these diagrams transforms the one Pomeron exchange amplitude
 into the
\begin{equation}
\label{a2}
A(s,b)=i(1-e^{-\Omega(s,b)/2})\ ,
\end{equation}
where the opacity $\Omega(s,b)$ is given in terms of the one Pomeron exchange amplitude $\Omega(s,b)=-iA_{\rm BFKL}(s,b)$. Here $A(s,b)$ is the Fourier transform
\begin{equation}
A(s,b)=\frac 1{4\pi^2}\int d^2q ~e^{i\vec{q}\cdot\vec{b}}~\frac{A(s,q^2)}{s}\ .  
\end{equation}
where $q^2=|t|$.   Note that $A(s,b)$ satisfies  the unitarity equation
 \begin{equation}
\label{un1}
2\mbox{Im}A(Y,b)=|A(Y,b)|^2+G_{\rm inel}(Y,b)\ ,
\end{equation}
where $G_{\rm inel}$ denotes the total contribution of all the inelastic channels. Here, for convenience, we have replaced the variable $s$ with the rapidity $Y=$ln$(s/m^2_p)$ where $m_p$ is the mass of the proton.
It is seen from (\ref{a2}) that when $\Omega$ increases the amplitude $A(s,b)\to i$, that is $A(s,b)$ tends to the black disk limit.

\begin{figure}[h]
\includegraphics[trim=-4cm 6cm 0cm 13cm 0cm,scale=0.6]{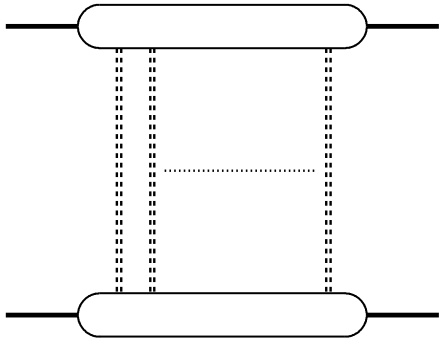}
\caption{\sf The multi-Pomeron exchange diagram which tames the asymptotic growth of the amplitude. Each double-dotted line corresponds to BFKL Pomeron exchange. }
\label{f1}
\end{figure}

Note that formally the absorptive effects are of the order of a $\alpha_s^2$ correction to the elastic amplitude. This correction can be large numerically for $\ln(s/s_0)\omega_0>2\ln(1/\alpha_s)$ due to the power growth of the BFKL amplitude proportional to $\alpha_s^2(s/s_0)^{\omega_0}$. However this  $\alpha_s^2$ correction to the amplitude cannot be considered as a next-to-next contribution to the BFKL kernel. This is a completely different correction.

The analogous absorptive corrections described by the multi-Pomeron diagrams are crucially important to restore  unitarity for processes with Large Rapidity Gaps (LRG)
where some groups of particles are separated from each other by large rapidity intervals. Indeed,
it was recognized already in the 1960s~\cite{VK-T,FK} that the multi-Reggeon reactions, 
\begin{equation}
pp\to p+X_1+X_2+...+X_n+p,
\label{eq:XXX}
\end{equation}
shown in Fig.2 may cause a problem 
with unitarity. Being summed over $n$ and integrated over the rapidities 
of each group, the cross section of such quasi-diffractive production increases 
faster than a power of $s$. This was  termed  in the literature
as the  Finkelstein-Kajantie (FK) disease,
see \cite{Abarbanel:1975me} for a review.
\begin{figure}
\vspace{-5cm}
\hspace{3cm}
\includegraphics[width=10cm]{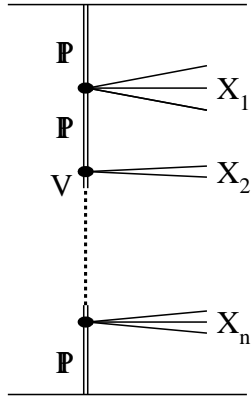}
\vspace{-1cm}
\caption{\sf The multi-Reggeon production process of eq.~(\ref{eq:XXX}).}.
\end{figure}


\begin{figure}[hbt]
\hspace{-1cm}
\vspace{-1cm}
\includegraphics[width=10.5cm]{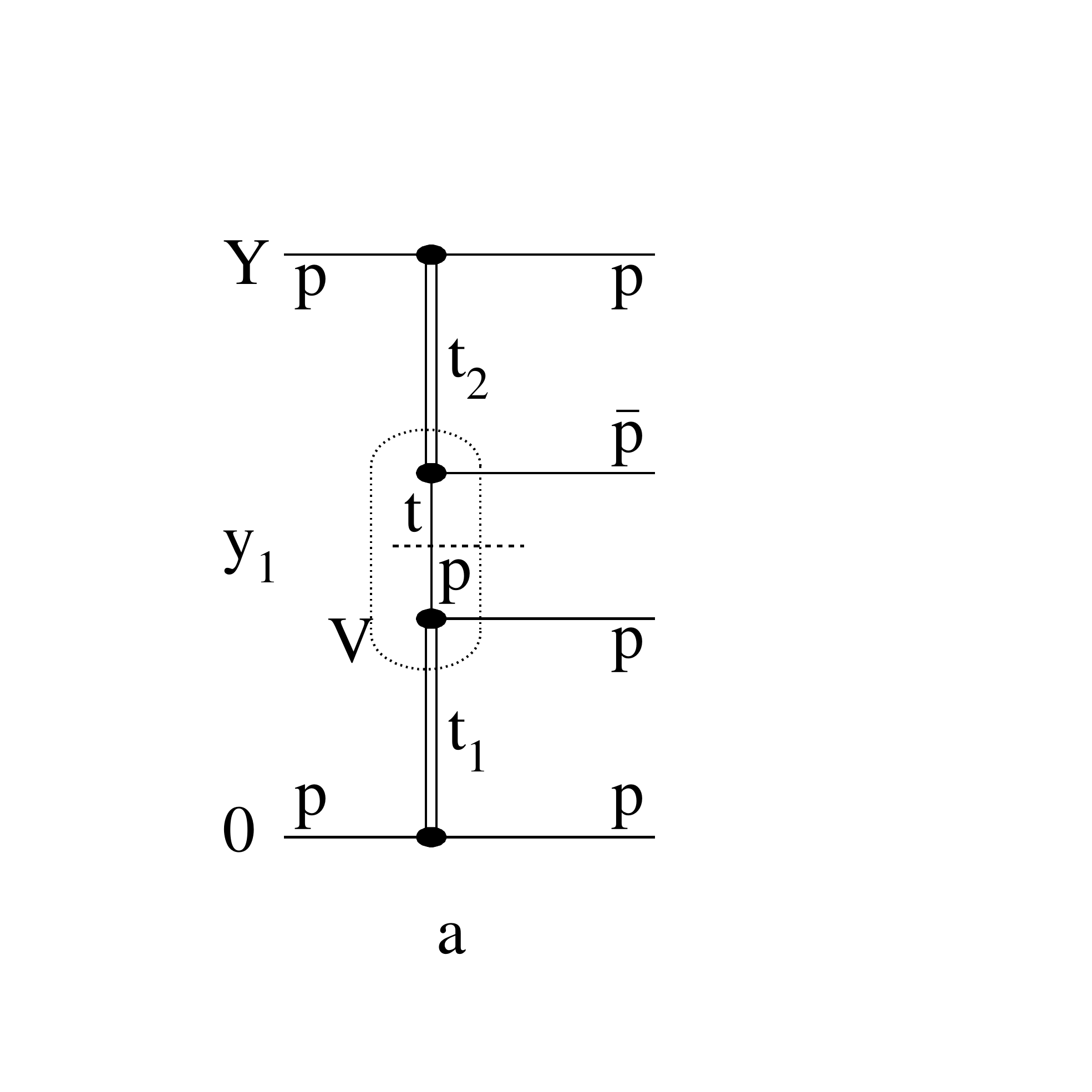}
\hspace{-3.5cm}
\includegraphics[width=10.5cm]{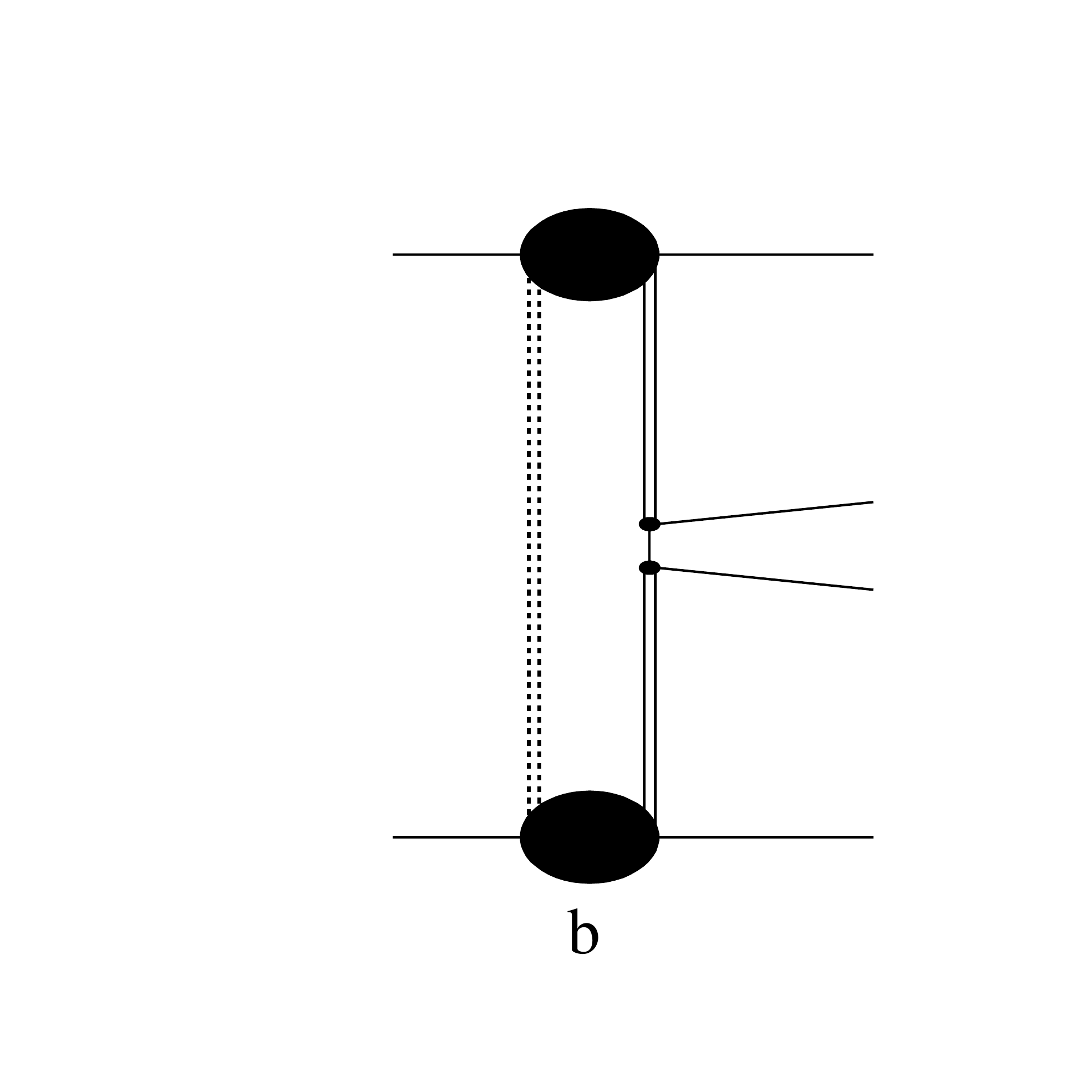}
\caption{\sf Proton-antiproton pair central exclusive  production}.
\end{figure}
Let us explain the problem using the simple example of Central Exclusive 
Production (CEP) of proton-antiproton pair, as shown in Fig.3a. Since the  proton-proton elastic cross section does not vanish but increases with energy,
 the corresponding contribution to inelastic cross section reads 
\begin{equation}
\label{t-int}
\sigma^{\rm CEP}~=~N\int_0^Y dy_1\int dt_1 dt_2 
|A(y_1,t_1)\cdot V\cdot A(Y-y_1,t_2)|^2\propto \int dy_1 \sigma_{el}(y_1)\sigma_{el}(Y-y_1)\ ,
\end{equation}
where the elastic amplitude $A(y,t)$ is normalized in such a way that  $\int dt |A(y,t)|^2=\sigma_{\rm el}(y)$, and where the upper rapidity $Y=\ln s/m^2_p$. In this example the vertex $V$ describes the central production of the $p\bar p$-pair. Thus we find
\be
\label{cep1}
\sigma^{\rm CEP}\propto \ln s\cdot\sigma_{\rm el}^2.
\ee 
 For a black (or grey) disk of increasing radius $R=c\ln s$,  this leads to
 \be
 \sigma^{\rm CEP}\propto (\ln s)^5~~~\gg ~~~\sigma_{\rm tot}\propto(\ln s)^2
 \ee

The same result can be obtained in impact parameter, $b$, space (see~\cite{black} for details). Moreover
 working in $b$ space we have a stronger constraint since for each value of 
$b$, that is for each partial wave $l=b\sqrt s/2$ of the incoming
 proton pair, we have the unitarity equation (\ref{un1}) 
and  the `total' cross section, $\sigma(b)_{\rm tot}$ must be less than the 
 corresponding CEP contribution . 

Actually one will face this FK problem in any model where the elastic cross section does 
not decrease with energy.

\section{Black disk solution of the FK problem}
The only known solution of this multi-Reggeon problem comes from `black disk' asymptotics 
of the high-energy cross sections. In such a case the (gap) survival probability, $S^2$, 
of the events with LRGs tends to zero as $s\to\infty$ and the
value of $\sigma^{\rm CEP}$ does not exceed $\sigma_{\rm tot}$. 

In other words besides the contribution of Fig.3a we have to consider the diagram of 
Fig.3b where the double dotted line denotes an additional proton-proton 
interaction. This diagram describes the absorptive correction to the original CEP process 
and has a negative sign with respect to the 
amplitude $A^{(a)}$ of Fig.3a. Therefore to calculate the CEP cross section we have 
to square the full amplitude
\begin{equation}
\label{ful}
|A_{\rm full}(b)|^2~=~|A^{(a)}(b)-A^{(b)}|^2~=~S^2(b)\cdot |A^{(a)}(b)|^2\ ,
\end{equation}
 where
 \be
 \label{10}
 S^2(b)=|e^{-\Omega(b)}|\ , ~~~~~{\rm with} ~~~~~{\rm Re}\Omega \ge 0 \ ,
 \ee
and $\Omega(b)$ is the opacity of incoming protons.

 Indeed, in terms of {\bf S}-matrix, the elastic component $S_l=1+iA(b)$; and the unitarity equation (\ref{un1})
reflects the probability conservation condition 
\be
\sum_n{\bf S}^*_l|n\rangle\langle n|{\bf S}_l~=~1
\ee
for the partial wave $l=b\sqrt s/2$. The 
solution of unitarity equation (\ref{un1}) reads
\begin{equation}
\label{el}
A(b)=i(1-e^{-\Omega(b)/2})\ ,
\end{equation}
or in terms of the partial wave amplitude with orbital moment $l=b\sqrt s/2$
\be
a_l~=~i(1-e^{2i\delta_l})~=~i(1-\eta_le^{2i{\rm Re}\delta_l})
\ee
where 
\be
\label{eta}
\eta_l=e^{-2{\rm Im}\delta_l}~~{\rm with}~~~ 0\le \eta_l \le 1.
\ee
The above discussion shows that $-\Omega(b)/2$ plays the role of $2i\delta_l$. The elastic component of {\bf S} matrix  $S_l=\exp(2i\delta_l)=\eta_l\exp(2i\mbox{Re}\delta_l)$.

 The gap survival factor $S^2$ is the probability to observe a pure CEP event where the 
 LRGs are not populated by secondaries 
 produced in an additional inelastic interaction shown by the 
 dotted line in Fig.3b. Thus according to (\ref{eta})
 \begin{equation}
 \label{gap1}
 |S(b)|^2~=~1-G_{\rm inel}(b)~=~\eta^2=~e^{-{\rm Re}\Omega(b)}\ .
 \end{equation}
 Equation (\ref{gap1}) can be rewritten as (see (\ref{el},\ref{eta}))
 \begin{equation}
 \label{gap2}
 |S(b)|^2=|1+iA(b)|^2=|S_l|^2\ .
 \end{equation}
 
 In the case of {\it black disk} asymptotics\footnote{Recall that the word `black' means the {\em complete} absorption of the incoming state (up to power of $s$ suppressed corrections), i.e. Re$\Omega(s,b)\to \infty$. `Black disk' means  
 that in some region of impact parameter space, $b < R$, the whole initial 
wave function is absorbed.
  That is, the value of $S(b) = 1 + iA(b) = S_ l \to 0$, i.e. $A(b) \to i$.}
 \be 
 {\rm Re}\Omega(b)\to\infty ~~{\rm and}~~ A(b)\to i,
 \label{eq18}
 \ee
 for $b<R$.   That is, we get $S^2(b)\to 0$. The decrease of the gap survival probability 
 $S^2$ overcompensates the growth of the original CEP cross section (Fig.3a), so that 
 finally we have no problem with unitarity.
 
 Recall that this solution of the FK problem was actually realized by Cardy 
 in~\cite{Cardy}, where the reggeon diagrams (generated by Pomerons with intercept 
 $\alpha_P(0)>1$) were summed up by {\em assuming analyticity} in the number of Pomerons in a multi-Pomeron  vertex. It was shown that the corresponding absorptive corrections  (analogous to that shown in Fig.3b) suppress not only the  growth of  a simplest,
 diagram Fig.3a, contribution but the growth of cross sections of  processes with  an arbitrary number of LRG, Fig.2 (for a more detailed discussion see ~\cite{black}). 

  Note that at the moment we deal with a one-channel eikonal.  
  In other words in Fig.3b and in the unitarity equation (\ref{un1}) we only account for the pure 
  elastic intermediate states 
  (that is the proton, for the case of $pp$ collisions). In general, there may be 
  $p\to N^*$ excitations shown by the black blobs in Fig.3b.
  The possibility of such  excitations can be included via the  
  Good-Walker formalism~\cite{GW} in terms of G-W eigenstates, $|\phi_i\rangle$,  which 
  diagonalize   the high energy scattering process; that is 
  $\langle\phi_k|A|\phi_i\rangle=A_k\delta_{ki}$. 
  In this case we encounter the FK problem for each state $|\phi_i\rangle $, and 
  then solve 
  it for the individual eigenstates.

Recall that the mean gap survival factor $\langle S^2\rangle$ strongly depends on the particular process. Since the opacity (the optical density) of the proton depends on impact parameter $b$ we expect a much larger $S^2$ for the reactions mediated by photons, such as ultraperipheral vector meson production or lepton pair production. These  processes occur at a large distances, $b$, where $\Omega(b)$ is already small. Correspondingly the survival factor 
$S^2(b)=|\exp (-\Omega(b))|$
 of (\ref{10}) is close to 1. On the other hand, the QCD induced exclusive production of a heavy object comes from a small distances, and is strongly suppressed by $S^2$. At small impact parameter the probability of an additional soft interaction, which fills the rapidity gap, is quite large. That is, there is a small chance for a large rapidity gap to survive.

 A more detailed description of the typical values and the behaviour of gap survival factors for different processes can be found in the review given in~\cite{KMR-gap}.

\section{Enhanced diagrams}  
  At first sight it looks sufficient to screen not the whole CEP amplitude, as in Fig.3b, but just the central vertex $V$ as in Fig.4a. 
  \begin{figure}
\vspace{-2.5cm}
\includegraphics[width=10.5cm]{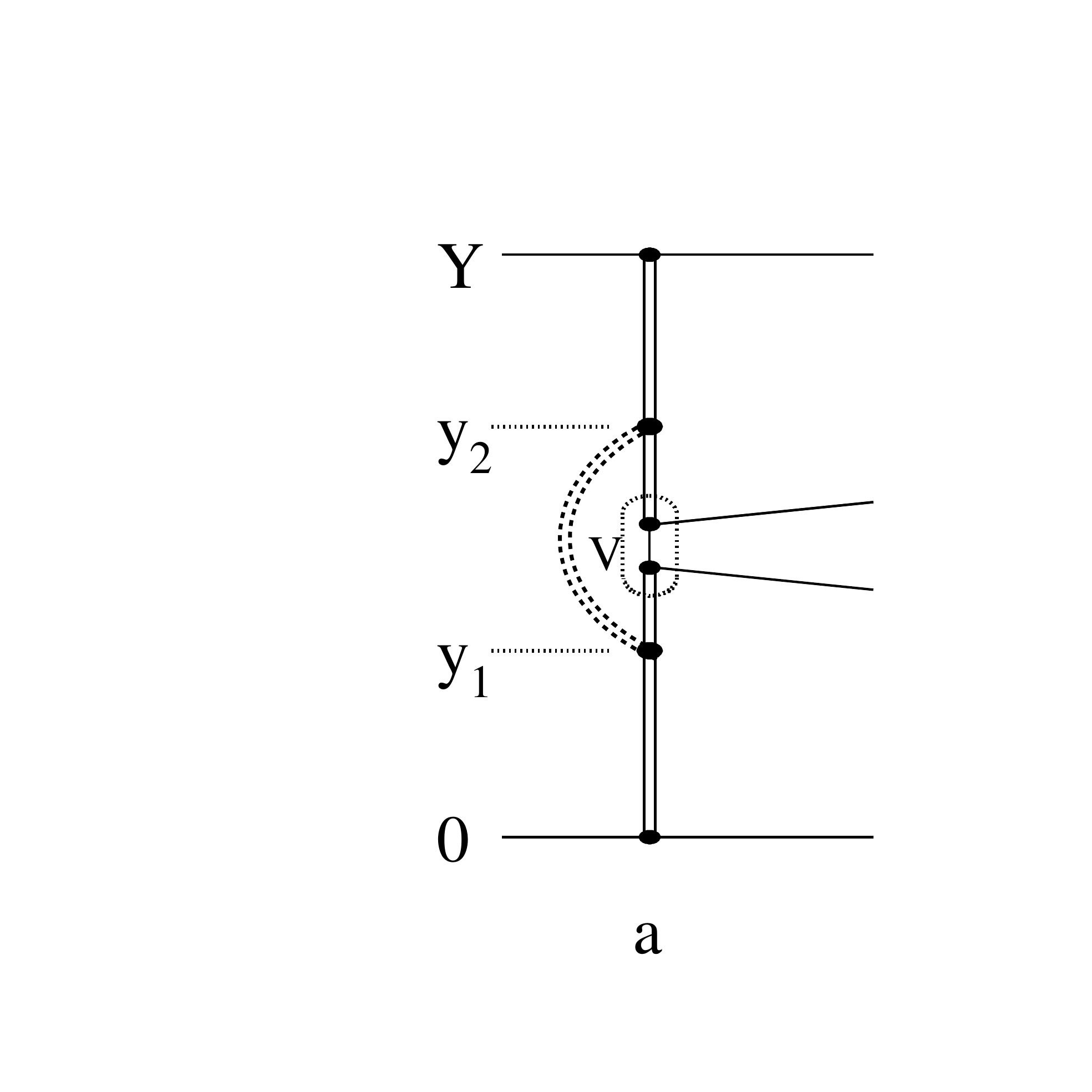}
\hspace{-3.5cm}
\includegraphics[width=10.5cm]{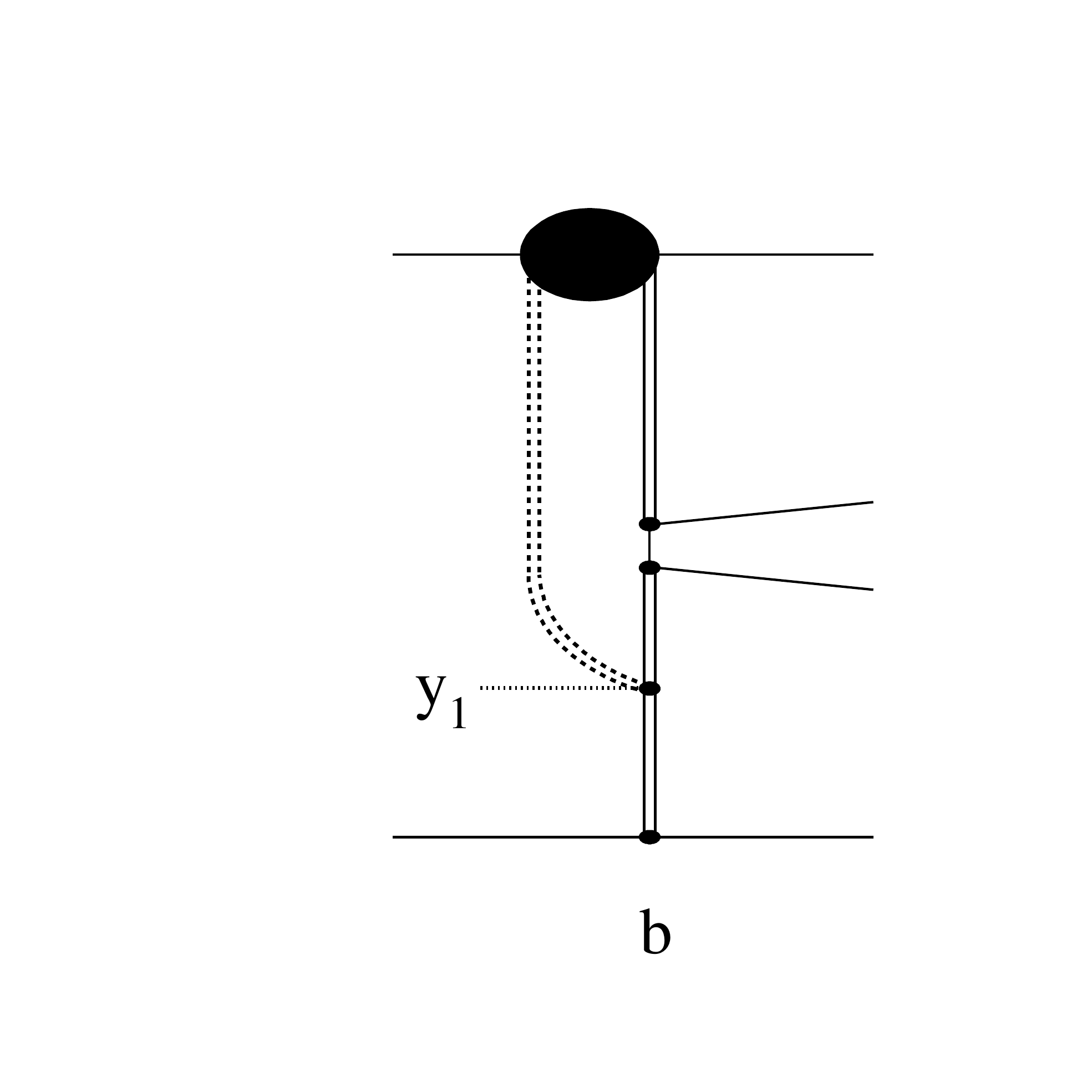}
\vspace{-2cm}
\caption{\sf Proton-antiproton pair central exclusive  production screened by enhanced (a) and semi-enhanced (b) diagrams.}.
\end{figure}
  Let us consider the enhanced diagram Fig.4a in more detail. Note that we have to integrate over the rapidity-positions of the `effective' triple-Pomeron vertices at $y_1$ and $y_2$. Since the original amplitude (shown by the double dotted  line) increases with energy, that is with the size of $|y_2-y_1|$ interval, the main contribution comes from the configurations where $y_1\to 0$ and $y_2\to Y$. In other words the enhanced diagram\footnote{This contribution is {\em enhanced} by the large intervals of integrations over $y_1$ and $y_2$.}, Fig.4a, diagram acts as the non-enhanced Fig.3b  graph considered above. The same is valid for the semi-enhanced diagram of Fig.4b.
  
  Still for the expected value of $\omega_0\simeq 0.2~-~0.3$
  the interval of integration ($|y_2-y_1|\sim 1/\omega_0\sim 4$) is not small  and the effect of enhanced screening is not negligible. Partly it can be described in terms of the Good-Walker eigenstates $|\phi_i\rangle$, but then we have to account for a large number of eigenstates, up to the mass
  \be
  M_i~\sim ~m_p\cdot\exp(0.5/\omega_0)~\sim ~5-10~{\rm  GeV}.
  \ee
  
Note however that the interval of $y_{1,2}$ integration actually depends on the mass of the centrally produced system and the scale corresponding to the subprocess described by the central vertex $V$. Since we are working in the low-$x$ region for a simplified rough  estimate we consider the double logarithmic DGLAP evolution~\footnote{The double logarithmic term is exactly the same for BFKL and DGLAP.} and take $\omega_0=0.3$ for the input distribution and account for the $(1-x)^5$ large $x$  behaviour of the gluon at the input scale $Q_0=1$ GeV. Note that the parton transverse momenta, $k_t$, are strongly ordered during the DGLAP evolution. The absorptive cross section 
$\sigma^{\rm abs}\propto 1/k^2_t$ decreases with $k_t$. Thus actually only input partons with  low  transverse momenta ($k_t <Q_0)$ participate 
in an additional interaction and so must be accounted for when we calculate the gap survival factor $S^2$.

Within the double logarithmic (DL) approximation the density of partons with momentum fraction $x$ at scale $\mu$ is given by the convolution of the input distribution $(1-x)^5x^{-\omega_0}$ with the DL exponent 
\begin{equation}
\label{y0}
\int_x^1 \frac{dx_1}{x_1}(1-x_1)^5x_1^{-\omega_0}\exp(\sqrt{16N_c/\beta_0\ln(x_1/x)\ln(G)})\ ,
\end{equation}
  where $G=\ln(\mu^2/\Lambda^2_{\rm QCD})/\ln(Q^2_0/\Lambda^2_{\rm QCD})$.

 In the case of a semi-enhanced diagram Fig.4b we get a similar expression but multiplied by the factor $x_1^{-\omega_0}$ coming from the screening Pomeron (shown by the double-dotted line in Fig.4b). Now the momentum fraction 
$x_1=\exp(-y_1)$ corresponds to the lower triple-Pomeron vertex in Fig.4b.

To obtain a more or less reasonable numerical factor,  we include in the integrand the ratio of the triple-Pomeron vertex, $g_{3P}$ to the elastic 
proton-Pomeron vertex, $g_N$, and assume that in  a semi-enhanced diagram the $t$-slope, $B_{\rm enh}$, corresponding to the Pomeron loop is twice smaller than that, $B_{\rm el}$, for  the case of a non-enhanced diagram Fig.3b. That is  
\begin{equation}
\label{y1}
\int_x^1 \frac{g_{3P}}{\pi g_N}\frac{B_{\rm el}}{B_{\rm enh}} \frac{dx_1}{x_1}(1-x_1)^5\exp(\sqrt{16N_c/\beta_0\ln(x_1/x)\ln(G)})\ ,
\end{equation}
In this way the ratio of the integrals (\ref{y1}) and (\ref{y0}) will be equal to the ratio of semi-enhanced to non-enhanced contributions.
  \begin{figure} 
\hspace{-.1cm}
\vspace{-.5cm}
\includegraphics[trim=-4cm 2cm 5cm 11cm,scale=0.6]{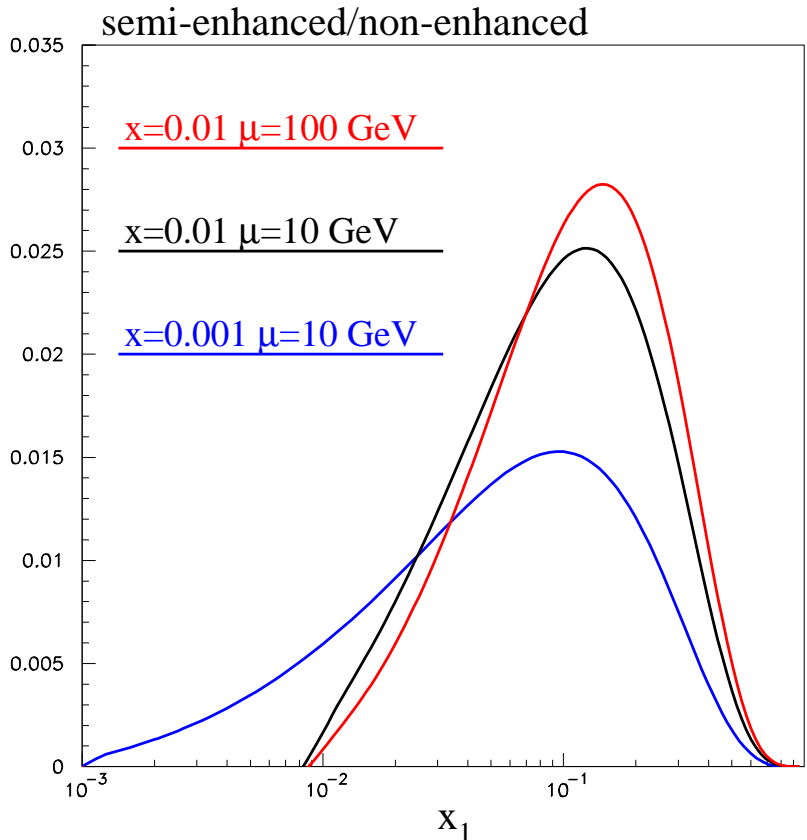}\vspace{1cm}
\caption{\sf The typical rapidity, $y_1$, distribution of the semi-enhanced, Fig.4b, absorptive correction; i.e the $x_1$ dependence of the integrand of eq.(\ref{y1}).} 
\end{figure}
  In Fig.5 we show the $x_1$ dependence of the integrand of (\ref{y1}) divided by the integral (\ref{y0}) for the case of a typical $x=0.01$ and scales $\mu=100$ and 10 GeV, and also for $x=10^{-3}$ and $\mu=10$ GeV; here we take the one loop running QCD coupling
  \be 
  \alpha_s(\mu^2)=4\pi/(\beta_0\ln(\mu^2/\Lambda^2) ~~~~~{\rm with} ~~\Lambda=200 {\rm MeV~~and~~}  \beta_0=8.333,
  \ee
  corresponding to 4 light quarks. Also we 
  take $\lambda=g_{3P}/g_N=0.2$~\cite{Luna} and $B_{\rm el}/B_{\rm enh}=2$.
    
Note that this way we overestimate the enhanced contribution, since:\\ 
  (a) actually $B_{\rm enh}>B_{\rm el}/2$ due to non-zero slope, $\alpha'_P>0$, of the effective Pomeron trajectory and non-zero size of the triple-Pomeron vertex.\\
  (b) in the non-enhanced case we only account for the elastic, $p\to p$ intermediate state, and neglect the proton, $p\to N^*$ excitations.
  
  It is seen that for a smaller final $x$ the maximum of integrand occurs at a smaller $x_1$ while for a larger final scale $\mu$ the maximum moves to a larger $x_1$. Depending on the mass (i.e. scale) and the momentum fraction carried by the central object the major contribution of enhanced diagram arises from the region of $x_1\sim 0.03-0.3$ (i.e. $y_1\sim 
  3.5-1.2)$. For a heavier object the value of $y_1$ decreases and the enhanced contribution becomes indistinguishable 
  from the non-enhanced absorptive correction of Fig.3b \footnote{The fact that enhanced screening becomes negligible for CEP of a heavy object was discussed in \cite{new}}.
  
  Recall that the major part of proton $p\to N^*$ dissociation was already included into the non-enhanced contribution via the  Good-Walker (GW) eigenstates. If the excitations up to $M_X=3.4$ GeV~\footnote{$M_X=3.4$ GeV is the value used by TOTEM collaboration to separate proton dissociation into  low- and high-mass states~\cite{TOT-low}.} are described within the GW formalism then only the region of $x_1\le 0.03$ should be described in terms  of enhanced diagrams.~\footnote{In this estimate we assume the mean transverse momentum $p_t\simeq 0.6$ GeV; $\;$ $ M_X^2=m_p^2+p^2_t/x_1$.} That is the absorptive corrections induced by enhanced Pomeron diagrams are practically negligible ($\sim O(0.01)$) for a large mass CEP in comparison with the non-enhanced effect.

\section{Summary} 
We emphasize that for the BFKL Pomeron, for which cross sections grow as the power $s^\omega$ of energy, black disk  absorption is the only cure of the FK disease. Otherwise we have the unphysical situation in which the cross section of events with Large Rapidity Gaps exceed the total cross section. Multi-Pomeron exchange diagrams account for the absorptive effect and compensate the growth of the cross section of the LRG events. The physical meaning of this phenomenon  is that simultaneously with the exclusive (LRG) process an additional inelastic interaction takes place, and secondaries from this additional interaction fill the rapidity gaps. Thus instead of an exclusive LRG event we get ordinary multiparticle inelastic production. The probability not to fill the gap is called the gap survival factor $S^2$.

At asymptotically high energies the value of $S^2\to 0$. That is the disc becomes black.
Any asymptotic behaviour with increasing high-energy cross section
 which does not lead to complete absorption is not consistent with multi-particle unitarity.
 
 We consider the role of absorptive corrections induced both by the enhanced and the un-enhanced diagrams and show that for the case of central exclusive production of a very heavy object the enhanced contribution act very similar to that from the non-enhanced diagrams. We emphasize that the value of $S^2$ is not universal but depends on the nature and the kinematics of a particular process.

 \section*{Acknowledgements}
 
VAK acknowledges  support from a Royal Society of Edinburgh  Auber award. MGR thanks the IPPP of Durham University for hospitality.


\thebibliography{ }
\bibitem{DGLAP}
  V.~N.~Gribov and L.~N.~Lipatov,
  Sov.\ J.\ Nucl.\ Phys.\  {\bf 15} (1972) 438
   [Yad.\ Fiz.\  {\bf 15} (1972) 781].\\
  L.~N.~Lipatov,
  Sov.\ J.\ Nucl.\ Phys.\  {\bf 20} (1975) 94
   [Yad.\ Fiz.\  {\bf 20} (1974) 181].\\
  G.~Altarelli and G.~Parisi,
  Nucl.\ Phys.\ B {\bf 126} (1977) 298.\\
  Y.~L.~Dokshitzer,
  Sov.\ Phys.\ JETP {\bf 46} (1977) 641
   [Zh.\ Eksp.\ Teor.\ Fiz.\  {\bf 73} (1977) 1216].
\bibitem{BFKL} 
V.~S.~Fadin, E.~A.~Kuraev and L.~N.~Lipatov,
  Phys.\ Lett.\  {\bf 60B}, 50 (1975);\\
E.~A.~Kuraev, L.~N.~Lipatov and V.~S.~Fadin,
  Sov.\ Phys.\ JETP {\bf 44}, 443 (1976)
  [Zh.\ Eksp.\ Teor.\ Fiz.\  {\bf 71}, 840 (1976)];\\
  E.~A.~Kuraev, L.~N.~Lipatov and V.~S.~Fadin,
  Sov.\ Phys.\ JETP {\bf 45}, 199 (1977)
  [Zh.\ Eksp.\ Teor.\ Fiz.\  {\bf 72}, 377 (1977)];\\
   I.~I.~Balitsky and L.~N.~Lipatov,
  Sov.\ J.\ Nucl.\ Phys.\  {\bf 28}, 822 (1978)
  [Yad.\ Fiz.\  {\bf 28}, 1597 (1978)].


\bibitem{NLL} 
V.~S.~Fadin and L.~N.~Lipatov,
  Phys.\ Lett.\ B {\bf 429}, 127 (1998)
  [hep-ph/9802290];\\
 M.~Ciafaloni and G.~Camici,
  Phys.\ Lett.\ B {\bf 430}, 349 (1998)
  [hep-ph/9803389];\\
  G.~P.~Salam,
  JHEP {\bf 9807}, 019 (1998)
  [hep-ph/9806482];\\
  M.~Ciafaloni and D.~Colferai,
  Phys.\ Lett.\ B {\bf 452}, 372 (1999)
  [hep-ph/9812366].
   



\bibitem{Fr}
Marcel Froissart,  Phys.Rev. 123 (1961) 1053-1057

\bibitem{VK-T} I.A.  Verdiev, O.V. Kancheli, S.G. Matinyan, A.M. Popova
and K.A. Ter-Martirosyan, Sov. Phys. JETP {\bf 19}, 1148 (1964).
\bibitem{FK}
J. Finkelstein, K. Kajantie,  Phys.Lett. {\bf 26B} (1968) 305-307.
\bibitem{Abarbanel:1975me}
   H.~D.~I.~Abarbanel, J.~B.~Bronzan, R.~L.~Sugar and A.~R.~White,
   Phys.\ Rept.\  {\bf 21}, 119 (1975).
\bibitem{Cardy}
  J.~L.~Cardy,
  Nucl.\ Phys.\ B {\bf 75} (1974) 413.
\bibitem{black}
V.A. Khoze, A.D. Martin, M.G. Ryskin
Phys.Lett. B780 (2018) 352-356,  arXiv:1801.07065 [hep-ph].
 \bibitem{GW} 
   M.~L.~Good and W.~D.~Walker,  
   Phys.\ Rev.\  {\bf 120} (1960) 1857.
\bibitem{KMR-gap} V.A. Khoze, A.D. Martin, M.G. Ryskin,
J.Phys. G45 (2018) no.5, 053002;
 arXiv:1710.11505.
 \bibitem{Luna}
  E.~G.~S.~Luna, V.~A.~Khoze, A.~D.~Martin and M.~G.~Ryskin,
  Eur.\ Phys.\ J.\ C {\bf 59}, 1 (2009)
  [arXiv:0807.4115 [hep-ph]].  
 
 \bibitem{new} V.A.~Khoze, A.D.~Martin and M.G.~Ryskin, JHEP {\bf 0605} (2006) 036  
 [arXiv:hep-ph/060224].
\bibitem{TOT-low} TOTEM collaboration, G. Antchev et al., EPL {\bf 101} (2013) 21003.
\end{document}